\documentclass[aps,pra,9pt,twocolumn,showpacs,superscriptaddress]{revtex4-2}
\usepackage{physics}							
\usepackage{latexsym}
\usepackage{amssymb}
\usepackage{graphics,epstopdf}
\usepackage{newlfont}
\usepackage{amsfonts}
\usepackage{epsfig}
\usepackage[colorlinks=true, citecolor=blue, urlcolor=blue]{hyperref}
\usepackage{amsmath}
\usepackage{graphicx}
\usepackage{dcolumn}
\usepackage{bm}
\usepackage{color}
\usepackage{longtable}
\usepackage{amsthm}

\newtheorem*{theorem*}{Theorem}

\begin{document}

\title{Concurrence speed limit and its connection with bounds in many body physics}

\author{Shrobona Bagchi}
\email{shrobona@kias.re.kr}
\affiliation{Quantum Universe Center, Korea Institute for Advanced Study, Seoul Campus, Korea.}

\begin{abstract} 
 Quantum speed limit is a fundamental speed limit for the evolution of quantum states. It is the single-most important interpretation of the time energy uncertainty relation. Recently the speed limit of quantum correlations have been proposed like the concurrence for pure quantum states. In this direction, we derive a speed limit bound for a quantum correlation named the concurrence for the generally mixed quantum states of two qubits. By this we mean that we find an expression for the minimum time required to reach a given value of entanglement starting from an arbitrary initial generally mixed state. We discuss the connection of the findings of this article in the interdisciplinary area of the condensed matter physics or the many body physics and quantum information science such as on the topic of Lieb-Robinson bound in a quantitative manner.
\end{abstract}

\maketitle

\section {Introduction }\label{intro}
In quantum theory, the uncertainty principle is one of the cornerstone results that differentiates it from the classical mechanics \cite{1,2,3}. Among the various type of uncertainty relations, the energy time uncertainty relation is one of the most important one \cite{4}. One of the consequences of the energy time uncertainty relation is the quantum speed limit \cite{5,6,7,8,9}. The quantum speed limit provides the bound for the minimum time required for a quantum state to evolve from a given initial state to a final state \cite{5,6,7,8,9}. Determination of quantum speed limit (QSL) is important in the areas of quantum metrology \cite{8}, quantum control \cite{8}, quantum thermodynamics \cite{8}, and quantum control \cite{8}. For instances, it plays a crucial role in determining the minimum charging time of quantum batteries \cite{8} and finding the minimum time required to implement quantum gates in quantum computation \cite{8}. Recently, the reverse QSL has been derived using the geometry of the quantum state space, and its application in the quantum battery was also discussed \cite{10}. The study of quantum speed limit for concurrence which is an entanglement measure is relevant for both fundamental and applied aspects, given the rapid progress in the area of quantum technologies and need to control the dynamics of quantum systems \cite{11}.

In another direction of study and research, the many body-physics is extremely important for the development of quantum technology since the quantum devices in general are systems of many bodies that have manifestly quantum effects \cite{12}. As a result the properties of many body physics will naturally come into play while implementing the quantum devices for advancing quantum technology \cite{12}. There has been considerable effort therefore to unravel the advantages the quantum physics of many body systems has to offer to advance quantum technology \cite{13}. In many body physics a bound named the Lieb-Robinson bound is very well known and has some similarity with the quantum speed limit bound \cite{14}. Both of these bounds are very relevant in the field of quantum technology. A connection between these two bounds will therefore prove to be fruitful in the area of quantum technology \cite{8}. Quantitative connections between these two bounds is yet to be done \cite{8}. Questions related to the connection between the Lieb-Robinson bound and the quantum speed limit are open to exploration and quantitative as well as qualitative characterization \cite{8}. This direction is open for more development and novel research. 

Our article is motivated by the above two bounds and the open area of research that connects the two together as well. The bound on the quantum speed limit derived here is useful in itself as a method to estimate the time taken to generate a certain amount of entanglement for general two qubit mixed quantum states in an analytical, simple and easy way. This result will be useful in the area of quantum information theory. Since we derived the bound for the correlation function specifically of entanglement measure called the concurrence, therefore we explore the topic of its connection with the Lieb-Robinson bound in a both the qualitative and quantitative fashion. We use the relation between the concurrence and the bipartite connected correlation function to arrive at our conclusion on this topic \cite{15}. We believe that this analysis can be extended further in the field of quantum information and computation and it will definitely be useful in the area of resource and time optimization for a more efficient quantum technological device.

In this paper, we find out the bounds for quantum speed limit of generating an entanglement value measured by the concurrence, starting from an initial value of entanglement measured by concurrence for the case of two qubits. We find these bounds for the case of pure states, mixed states, and unitary evolution. We keep our focus only on one entanglement measure called the concurrence. We start with the required background in section II. Then we move onto our results in section III. In this section III, we derive results for the case of pure states for all dimensions. In next section IV, we lay out different methods for the calculation of this quantum speed limit bounds for the case of qubits and mixed states. In section V, we derive quantum speed limits for the generation of entanglement captured by the generalized concurrence for higher dimensions for the case of mixed states. In section VI, we derive bounds for the multiparty case in which case the genuine entanglement measure is given by the geometric mean of generalized concurrence of the bipartitions. In section VII, we discuss a few applications in control theory where our analysis can be useful. In section VIII we end with conclusions and future directions.

\section {Background }\label{background}

Here in this section we briefly discuss some necessary background. These background include the time-energy uncertainty relation, definition and description of quantum speed limit, definitions of concurrence, generalized concurrence, properties of unitary operators in terms of stochastic matrices, unitary orbits of mixed quantum states, parametrization of quantum states with the same concurrence and the parametrization of the $SL(2,C)$ and $SL(2,R)$ matrices.

\subsection {Energy Time Uncertainty Relation}

The topics needed for our analysis include the energy time uncertainty relation. In his original paper \cite{1}, Heisenberg began by deriving the uncertainty relation for position and momentum on the basis of a supposed experiment in which an electron is observed using a $\gamma$-ray microscope. By consideration of the theory of the Compton effect he could argue that the precision of the determination of position x and momentum p are connected by the uncertainty relation. 
Heisenberg also discusses the classically conjugate variables of time and energy and defines a time operator through the, quote, ``familiar relation” as follows 
\begin{align}
[E,t]=-i\hbar
\end{align}
and on the basis of this assumes a time energy uncertainty relation $\Delta E\delta t~\hbar$. Later Mandelstam and Tamm realized that the energy time uncertainty relation is not about simulatneous measurements \cite{7}. However it is about the intrinsic time scale of unitary dynamics. This means that one should write $\Delta t> \frac{\hbar}{\Delta E}$ where one then interprets $\Delta t$ as  the time a quantum system needs to evolve from an initial to a final state \cite{7}.

\subsection {Quantum Speed Limit}

In particular Mandelstam and Tamm derived the first expression of the quantum speed limit time $\tau_{QSL} = \frac{\pi}{2\Delta H}$, where $\Delta H$ is the variance of the Hamiltonian, $H$, of the quantum system \cite{7}. As an application of their bound, they also argued that $\tau_{QSL}$ naturally quantifies the life time of quantum states, which has found widespread prominence in the literature \cite{8}. With the advent of quantum computing Mandelstam and Tamm’s interpretation of the quantum speed limit as intrinsic time-scale experienced renewed prominence. Their interpretation was further solidified by Margolus and Levitin \cite{9}, who derived an alternative expression for $\tau_{QSL}$ in terms of the (positive) expectation
value of the Hamiltonian, $\tau_{QSL} = \frac{\pi}{2\langle H\rangle}$. Eventually, it was also shown that the combined bound,
\begin{align}
\tau_{QSL}=\max\{\frac{\pi\hbar}{2\Delta H},\frac{\pi\hbar}{2\langle H\rangle}\}
\end{align}
is tight. This means it sets the fastest attainable time-scale over which a quantum system can evolve. In particular, it sets the maximal rate with which quantum information can be communicated \cite{16}, the maximal rate with which quantum information can be processed \cite{17}, the maximal rate of quantum entropy production \cite{18,19}. It also has very important applications in quantum metrology and quantum control theory \cite{20,21,22}.

\subsection {Concurrence and generalized concurrence}

The concurrence is an entanglement monotone defined for a mixed state of two qubits as:
\begin{align}
C(\rho)=\max\{0,\lambda_1-\lambda_2-\lambda_3-\lambda_4\}
\end{align}
where, $\{\lambda_i\}$ are the eiegnvalues, arranged in decreasing order of the following Hermitian matrix
\begin{align}
R=\sqrt{\sqrt{\rho}\tilde{\rho}\sqrt{\rho}}
\end{align}.
with $\tilde{\rho}=(\sigma_y\otimes\sigma_y)\rho^*(\sigma_y\otimes\sigma_y)$, the spin-flipped state of 
$\rho$  and $\sigma_y$ a Pauli spin matrix. The complex conjugation $*$ is taken in the eigenbasis of the Pauli matrix $\sigma _{z}$. A generalized version of concurrence for multiparticle pure states in arbitrary dimensions is defined as:
\begin{align}
C(\rho)=\sqrt{2(1-(\mathrm{Tr}(\rho_M)^2))}
\end{align}
A regularized expression, we adopt in this paper, can be written as
\begin{align}
C(\rho)=\sqrt{\frac{d_{min}}{d_{min}-1}(1-(\mathrm{Tr}(\rho_M)^2))}
\end{align}
where $d_{min}$ denotes the dimension of the smaller subsystem. It is well known that a particle cannot freely share entanglement with two or more particles. This restriction is generally called monogamy. However the formal quantification of such restriction is only known for some measures of entanglement and for two-level systems. The first and broadly known monogamy relation was established by Coffman, Kundu, and Wootters for the square of the concurrence \cite{23}.

\subsection {Birkhoff's theorem for Unitary operators and stochastic matrices}

The Birkhoff’s theorem states that any doubly stochastic matrix lies inside a convex polytope with the permutation matrices at the corners. This is a well known result in mathematics. It can be proven that a similar theorem holds for unitary matrices with equal line sums for prime dimensions. In the present paper, we use an equivalent of Birkhoff’s theorem for unitary matrices for our purpose of finding concurrence speed limit for general mixed states for two qubits. The importance of unitary matrices equally follows from physics, more in particular from quantum physics and quantum information. In contrast to the $n \otimes n$ doubly stochastic matrices, the $n \otimes n$ unitary matrices form a genuine group, called the unitary group and denoted $U(n)$. Within this group figures a subgroup denoted $XU(n)$: the group of $n \otimes n$  unitary matrices with all row sums and all column sums equal unity. It is also well known that, $XU(n)$ acts as a ‘doubly stochastic’ analogon within $U(n)$.

\subsection {Unitary orbits of mixed states}

The condition of a closed quantum system imposes the condition of unitarity on the evolution of the quantum state in the driving for the calculation of the fixed concurrence speed limit. The constraint of unitary evolution further imposes constraints on the powers of trace of the initial, final and intermediate quantum states while the driving Hamiltonian is switched on. These constraints have been studied in the earlier literature, which are stated as the following equivalent conditions.  Let $\rho_1$ and $\rho_2$ be two density matrices. The following are equivalent:
\begin{itemize}
\item $\rho_1$ and $\rho_2$ are unitarily equivalent, i.e., $\rho_2 = U\rho_1U^\dagger$ for some unitary matrix $U$.
\item $\rho_1$ and $\rho_2$ have the same spectrum, i.e., the same eigenvalues including multiplicity.
\item $Tr(\rho_1^r)=Tr(\rho_2^r)$ for all $r=1,2,...,n$.
\end{itemize}
This result shows immediately that the orbit of a density matrix under $U(n)$ is uniquely determined by its spectrum, i.e., two density matrices belong to the same unitary orbit if and only if they have the same eigenvalues $\lambda_i$ with the same multiplicities $n_i$.

\subsection {Parametrization of quantum states with the same concurrence}

To this end we need a parameterization of the manifold of states with constant concurrence \cite{24}
\begin{align}
\rho'=\frac{(A\otimes B)\rho(A\otimes B)^\dagger }{\mathrm{Tr}((A\otimes B)\rho(A\otimes B)^\dagger )}
\end{align}
The transformation rule is:
\begin{align}
C(\rho')=C(\rho)\frac{|\mathrm{det}A||\mathrm{det}B|(A\otimes B)\rho(A\otimes B)^\dagger }{\mathrm{Tr}((A\otimes B)\rho(A\otimes B)^\dagger )}
\end{align}
It was furthermore shown that for each density matrix $\rho$ there exists an $A$ and $B$ such that $\rho'$ is Bell diagonal. The concurrence of a Bell diagonal state is only dependent on its largest eigenvalue
\begin{align}
\mathrm{Tr}((\frac{AA^\dagger}{|\mathrm{det}A|}\otimes(\frac{BB^\dagger}{|\mathrm{det}B|})\rho)=1
\end{align}
It is clear that we can restrict ourselves to matrices $A$ and $B$ having determinant 1 $(A, B \in SL(2, C))$, as will be done in the sequel.

\subsection {Parametrization of the $SL(2,C), SU(2)$ and $SL(2,R)$ matrices}
For the 2-dimensional $SU(2)$ matrices, there is a fairly general parametrization formulation which is as follows:
\begin{align}
s_2=\begin{bmatrix}
e^{i\alpha}\cos\theta & -e^{-i\beta}\sin\theta\\
e^{-i\beta}\sin\theta & e^{-i\alpha}\cos\theta
\end{bmatrix}
\end{align}
\begin{align}
s_2=\begin{bmatrix}
e^{i\alpha}\cos\theta & -e^{-i\beta}\sin\theta\\
e^{-i\beta}\sin\theta & e^{-i\alpha}\cos\theta
\end{bmatrix}
\end{align}

\subsection{Lieb-Robinson bound in many body physics}
The Lieb Robinson bound is a theoretical upper limit on the speed of propagation of information or correlation in a non-relativistic quantum system \cite{14}. This was first introduced and formulated by Elliott Lieb and Derek Robinson \cite{14}. The bound on the speed is called the Lieb-Robinson velocity. This bound has found relevance in quantum information theory, quantum computation, many body physics and condensed matter physics \cite{14}. There are a multitude of Lieb-Robinson bound that have been developed by various authors thereafter \cite{14}. In this article we refer to a Lieb-Robinson bound that gives an upper bound to the speed of propagation of the correlation function called the connected correlation function in a finite dimensional quantum system such as a lattice or a spin chain. We state here what this Lieb-Robinson mean for the connected correlation function. 

For any observables $A$ and $B$ in a finite dimensional Hilbert space with finite support $X\in \Gamma$ and $Y\in\Gamma$, respectively, for any time $t\in \mathbb{R}$, for some following positive constants $a,c,v$.
\begin{align}
||[A(t),B]||\leq exp(-a(d(X,Y)-v(t))),
\end{align}
where $d(X,Y)$ denotes the distance between the sets $X$ and $Y$. $v$ is called the group velocity or the Lieb-Robinson velocity. 
Many applications have been found for the Lieb-Robinson bound after its discovery.  Among those applications, the main applications are for example finding the error bounds \cite{14} and the formulation of the Lieb–Schultz–Mattis theorem \cite{25}. 

\section{Main Methods }

\subsubsection{The method to find the quantum speed limit for states with fixed concurrence for general two qubit states}

At first we choose a Bell diagonal state with the same concurrence which we want to achieve. After this we choose a specific distance measure that will give us the speed limit. We minimize this value for the given Bell state over all possible tensor product of $SU(2)$ for closed quantum systems. The calculation is simplified when the states are symmetric with respect to the two parties. In that case we have tensor product of the same copies of the unitary operators. There is two step minimization. At first we minimize with respect to the unitary operators. Then we do this for all the Bell states with the same concurrence. At the end we take the minimum from all the minimizations of Bell states. This gives us the correct speed limit to attain that value of entanglement from an initial value of entanglement. For open quantum systems, we do not take unitary operator. Instead if we consider only the Bloch plane, i.e., real Bloch vector, then we consider the $SL(2,R)$ matrices. We illustrate this with a few examples for symmetric states, asymmetric states, different initial values of entanglement , open and closed quantum systems and for different measures of distance. We also verify the bounds numerically with random quantum states. We take the initial quantum state with a fixed amount of entanglement as $\rho_i$ and we suppose we want to have a bound on the shortest time for it to evolve to a quantum state with $E$ amount of quantum entanglement captured by a value of the concurrence $C_E$. For this purpose, we first choose the suitable Bell state with $E$ value of the quantum entanglement. A Bell diagonal state is a 2-qubit state that is diagonal in the Bell basis. In other words, it is a mixture of the four Bell states. The form of the Bell diagonal state is therefore given by the following
\begin{align}\nonumber
\rho_{BD}=p_1|\Phi^+\rangle\langle\Phi^+|+p_2|\Phi^-\rangle\langle\Phi^-|\\ +p_3|\Psi^+\rangle\langle\Psi^+|+p_4|\Psi^-\rangle\langle\Psi^-|,
\end{align}
where, we have $\sum_ip_i=1$ and $|\Phi^+\rangle,|\Phi^-\rangle,|\Psi^+\rangle,|\Psi^-\rangle$ as the four Bell states in the two qubit Hilbert space.  The entanglement of the Bell diagonal state is given by the following
\begin{align}
C_{BD}=\max\{0,2\lambda-1\},
\end{align}
where, $\lambda=\max_i\{p_i\}$. Now we want $C_{E}=C_{BD}$. Therefore we fix the coefficients of the Bell diagonal state accordingly to fit this constraint. After this, we choose the distance measure as the  the trace distance $\mathrm{Tr}(\rho_i\rho_f)$. Now we have chosen $\rho_f=\rho_{BD}$ and for closed quantum systems subject to unitary driving the set of states having the same concurrence as $\rho_{BD}$ is given by $(U_A\otimes U_B)\rho_{BD}(U_A\otimes U_B)^\dagger$, where $U_A,U_B$ are unitary matrices belonging to the $SU(2)$ group. Only $SU(2)$ matrices suffice because let $\rho_i$ be the initial state of the driving Hamiltonian. Let $ \rho_{f1}$ be the first final state driven according to the Hamiltonian $H_1$. Therefore $\rho_{f1}=U_1\rho_i U_1^\dagger$. Similarly, $\rho_{f2}=U_2\rho_i U_2^\dagger$. As a result from these two relations, we see that $\rho_{f1}$ and $\rho_{f2}$ must be connected by unitary operators. Therefore, all set of final states must be connected by unitary operators. As a result we have to exclude all $SL(2,R)$ matrices which are not unitary in our consideration of the parametrization of the final quantum states having the fixed given value of concurrence. As a result, we consider only $SU(2)\otimes SU(2)$ matrices in our calculation. For open quantum systems, we consider the $SL(2,R)$ and $SL(2,C)$ matrices. As a first step of minimization, we minimize the trace distance over all such unitary matrices by using the parametrization of the $SU(2)$ matrices. After this we minimize over all such possible Bell states with the same amount of concurrence.

\subsubsection{Upper bound and lower bound on the minimum time taken to generate or degrade generalized concurrence for bounded Hamiltonians}

In the calculation of the quantum speed limit, we have the numerator that denotes the distance between the initial and the final quantum states and the denominator that is a time averaged standard deviation of the driving Hamiltonian from the initial state to a final quantum state for the case of unitary driving. For many cases of initial and final mixed states, an optimal Hamiltonian is not possible to find analytically and sometimes it does not even exist for the calculation of the quantum speed limit. In that case we take the Hamiltonians while standard deviation must lie within a range of predefined values, the Hamiltonians being called as the bounded Hamiltonians. These Hamiltonians can be realized in experimental scenarios, which we discuss in later sections of this paper. This has been discussed in the literature previously.

\subsubsection{Example 1: Closed quantum system and symmetric separable two qubit mixed state as initial state}

For the analysis of the first case, we consider the initial state as a separable state for two qubits which is diagonal in the computational basis. As a result it is a symmetric state with respect to the party A and party B. We take the initial state as $\sigma$ such as the following
\begin{align}
    \sigma=p_1|00\rangle\langle 00|+p_2|01\rangle\langle 01|+p_3|10\rangle\langle 10|\\ \nonumber
    +(1-p_1-p_2-p_3)|11\rangle\langle 11|
\end{align}
\vskip 10 pt 
where,$0\leq p_i\leq 1$. We want to calculate the optimal distances of this quantum state from the quantum state with a given or fixed value of quantum entanglement captured by concurrence for the two qubit states denoted by C. Based on this entanglement value we find out the quantum state which has the optimal distance from the initial quantum state. Now as we have stated above, we want to consider closed quantum systems and as a result consider only unitary operators for optimization. In this case we consider a symmetric state as the initial state and we also know that the final state is also a symmetric state with respect to parties A and B. We show below that this implies that it suffices to consider only unitary operators of the form $U \otimes U$ for the optimization. The matrix differential equation for the optimization of the quantity $\mathrm{Tr}(\rho\sigma)$ with respect to matrices $U_A$ and $U_B$ are given below as follows

\begin{align}
\frac{\partial}{\partial U_A}\otimes\mathbb{I}_2(\mathrm{Tr}(U_A\otimes U_B)\rho_{BD}(U_A^\dagger U_B^\dagger)\sigma)=0\\ \nonumber
\mathbb{I}_2\otimes\frac{\partial}{\partial U_B}(\mathrm{Tr}(U_A\otimes U_B)\rho_{BD}(U_A^\dagger U_B^\dagger)\sigma)=0
\end{align}
subject to the constraints $UAUA^\dagger = UA^\dagger UA = UBUB^\dagger = UB^\dagger UB = I_2$ for unitarity condition to be satisfied. Now if we trace out the first party from the first second equation and the second party from the first equation, we end up getting the exactly same equations of optimization with respect to the unitary operators above, since party A and B are completely symmetric in this case. As a result, we can conclude that for the optimization condition we have $UA = UB$ . Therefore going forward we can make the simplification that for the optimized state we have $\rho_0 = U \otimes U \rho B DU^\dagger \otimes U^\dagger $ and we can perform the explicit optimization over these subset of quantum states. In other words it is clear from the above equations that the parameters of the unitary operator that gives solutions to the first equation also gives the solutions of the second equation in exactly the same way. We now proceed with our explicit optimization procedure as follows. At first, we fix the initial quantum state with the choice of the parameters as
\begin{align}
    \sigma=p_1|00\rangle\langle 00|+p_2|01\rangle\langle 01|+p_3|10\rangle\langle 10|\\ \nonumber
    +(1-p_1-p_2-p_3)|11\rangle\langle 11|
\end{align}

We take the parametrization of the unitary operators as stated in the background section and calculate the trace as follows
\begin{align}\nonumber
\mathrm{Tr}(\rho\sigma)=T=\frac{1}{8}(2-2b3+e1+b2(1-2r2-2r3)+ \\ \nonumber 2(2b3-e1)(r2+r3)) + \\ \nonumber (-1+2r2+2r3)(-1((-2+3b2+2b3+3e1)\cos (4\gamma)+ \\ \nonumber 2(b2-e1)\cos(2(\alpha+\beta))\sin((2\gamma)^2)))
\end{align}
Now we have to optimize the above equation with respect to the parameters of the unitary operator namely $\alpha, \beta, \gamma $. For simplification of our calculation we take $sin(2\gamma)2 = t$ and $cos(\alpha + \beta) = s$ as single variables and perform our optimization with respect to them. After finding solutions for $s, t$, we can transfer back our solutions to the unitary operator parameters $\alpha, \beta, \gamma $. After finding the solutions to the first order differential equations, we will perform the second order differentiation to check which solutions give us the maximum distance and which solutions give us the minimum distance. The differential equations read as the following
\begin{align}
\frac{\partial T}{\partial b1}=\frac{\partial T}{\partial b2}=\frac{\partial T}{\partial b3}=\frac{\partial T}{\partial t1}=\frac{\partial T}{\partial t2}=0
\end{align}
which gives us the following solutions that the following are true simultaneously or separately
\begin{align}\nonumber
(-1+2r2+2r3)=-1+t1=b2-e1=\\ \nonumber 
(-2+2b3+3e1-e1t2+b2(3+t2)))=0
\end{align}
Now let us check which of these solutions give us the maximum or minimum based on the second derivatives of the trace function at those values and subsequently the corresponding Hessian matrix. Now from the above equations, if $t1 = 1$, then we definitely have $2r2 + 2r3 = 1$. Then we may not have other constraints on the other variables as 
$t1 = 1$ makes all other equations true trivially. Under these conditions the Hessian matrix is not semi-definite, i.e., its determinant gives us zero, as a result under those conditions, we cannot say if we have a maximum or minimum. Now, let us check the other solutions, for the cases where $2r2 + 2r3 = 0$ and $t1 = 1$.
In that case we directly see that 
$ b2 = e1 $ and 
$ b3=1-3e1$.
In this case, we can have any arbitrary value of $t2$. We find here that analytically we cannot find the exact solution by analysing second order differentials. In that case, we just find the values of the resulting equations and analyze if what we get is indeed the global minimum or maximum which is pretty accurate due to the fact that the function is simple as well as there are lowest number of variables involved. So we analyze individual cases, as follows one by one. At first, we note that the necessary condition for a maxima or a minima is that the first derivative should be zero. We find these satisfying conditions and plot the trace function at those values. An example is given by the following plot. From the plot we can easily calculate the maximum and the minimum values.

\subsection{Verification of our bounds numerically}

In this section, we verify that the bounds we have given in the previous sections are indeed valid. At first we verify the bounds for the symmetric initially separable two qubit states and then verify the bounds for the initially entangled asymmetric quantum state for the generation of a given entangled value. The first example we take is from the previous section.

\subsection{ Example 1: Symmetric initially separable quantum state}

In this example we choose the initial separable state of the form given by the following equation where we put the values of the parameters specifically so that we get the following trace function as follows
\begin{align}\nonumber
\frac{1}{8}(2-2b_3+0.75+b_2(1-20.2-20.1)\\ \nonumber +7\times 2(2b_3-0.75)(0.2+0.1)\\ \nonumber +(-1+20.2+20.1)(-((-2+3b_2+2b_3+30.75))))
\end{align}

\begin{figure}
\includegraphics[scale=0.3]{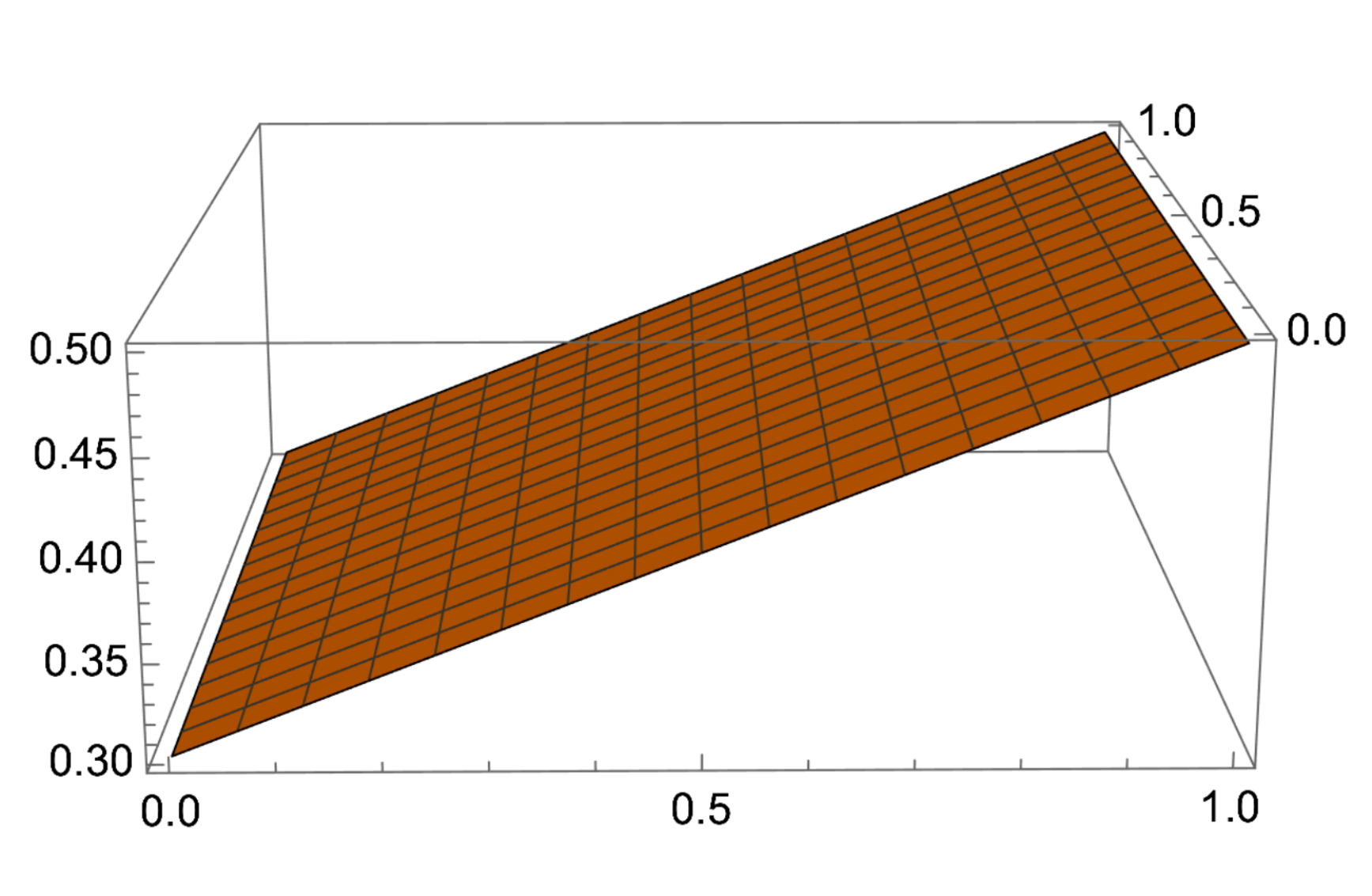}
\caption{Value of the trace function according to the example we have taken here. The vertical axix signifies the value of the trace function whereas the horizontal axes denote $b2,b3$.}
\end{figure}
The plot of the  function is given below. From this plot, it is easy to derive the maximum and the minimum values, from where we can derive the speed limit bounds using the value of the bounded Hamiltonians. For verifying the bound, we have to calculate the R.H.S. and L.H.S. separately and check if the inequality holds.

\section{An easily calculable lower bound on the distance for all sets of two qubit mixed quantum states with fixed concurrence values}
 
In general it is quite difficult to solve the optimization equation for an initially asymmetric quantum state, since now we have a larger number of variables to handle as the unitary matrices have to be taken different now. It will generally be a set of first order differential equations of polynomials of 8th order 6 variables that need to be solved simultaneously. We know from existing literature it is almost impossible to solve these high order polynomials for solutions in the general case. While for some cases, it might become simpler due to the particular choice of coefficients, in principle it can become very complicated to solve for arbitrary value of coefficients. As a result, we ask whether we can we give a bound or find an exact solution to our minimization of the trace condition analytically ? While it may be possible, there are no known solutions to the following minimization problem as far as we know
\begin{align}
\min_{U_A\otimes U_B}\mathrm{Tr}((U_A\otimes U_B)\rho(U_A^\dagger\otimes U_B^\dagger\sigma))
\end{align}
However it is possible to calculate an easily calculable lower bound on this analytically, which we describe hereafter. For this first let us note that $\rho$ and $\sigma$ are positive semidefinite Hermitian matrices which are diagonalizable by unitary operators. 
Let the unitary operator $U_\rho$ diagonalize $\rho$ and $U_\sigma$ diagonalize $\sigma$. 
We know the eigenvalues are positive semidefinite for density matrices. Then we denote $\rho_D$ and $\sigma_D$ as the corresponding diagonalized density matrices respectively. Therefore we have the following
\begin{align}\nonumber
\rho=U_\rho\rho_DU\rho^\dagger ~~~\sigma=U_\sigma\sigma_DU\sigma^\dagger\\ 
\rho_D=\sum_i(\lambda\rho)_i|i\rangle\langle i| ~~~\sigma_D=\sum_j(\lambda\sigma)_j|j\rangle\langle j| \nonumber
\end{align}
For any general quantum states for some fixed values of entanglement, the matrices $U_\rho$ and $U_\sigma$ are in general non-local for nonzero values of entanglement for initial and final quantum states. 
Let the eigenvalues be arranged in ascending order for $\rho$ and descending order for $\sigma$. Now for the bound we note the following
\begin{align}\nonumber 
\min_{U_A\otimes U_B}\mathrm{Tr}((U_A\otimes U_B)\rho(U_A^\dagger\otimes U_B^\dagger\sigma))\geq \\ \nonumber L_{C\sigma}=\min_U \min_{U}\mathrm{Tr}((U)\rho_D(U\dagger\sigma_D))
\end{align}
since the minimization now is over a larger set encompassing all possible unitary matrices in the two qubit space and where we have absorbed the diagonalizing unitary operators in the set of unitary matrices over which we are minimizing. We know that we can write the unitary operators as the following
\begin{align}
U=\sum_{ij} u_{ij}|i\rangle\langle j|
\end{align}
for the case of two qubits. Therefore, putting these value we compute the lower bound as the following
\begin{align}\nonumber
L_{c\sigma}=\min\mathrm{Tr}(\sum_{ij} u_{ij}|i\rangle\langle j|(\sum_k(\lambda\rho)_k|k\rangle\langle k|)\\ \nonumber(\sum_{lm} u_{lm}^*|m\rangle\langle l|)(\sum_n(\lambda\sigma)_n|n\rangle\langle n|) )\\ \nonumber
=\min(\sum_{ij=1}^2 |u_ij|^2(\lambda\rho)_i(\lambda\sigma)_j)
\end{align}
Now it is easy to see that U are unitary operators, therefore |uij|2 are elements of a doubly stochastic matrices, since the rows and columns of the unitary matrices are orthonormal vectors in complex Hilbert space. As a result, the above minimization can be written as the following minimization over doubly stochastic matrices
\begin{align}\nonumber
L_{c\sigma}=
\min(\sum_{ij=1}^2 |u_{ij}|^2(\lambda\rho)_i(\lambda\sigma)_j) \\ \nonumber \geq \min_S(\sum_{ij=1}^2 |s_{ij}|^2(\lambda\rho)_i(\lambda\sigma)_j)
\end{align}
where $S = {sij}$ are double stochastic matrices. Now, we apply the Birkhoff’s theorem for the doubly stochastic matrices. Let us observe that the quantity $P2i,j =1 sij (\lambda\rho )i (\lambda\rho )j$ is a linear function in $s_{ij}$ and the set of doubly
stochastic matrices is convex and compact. As a result, the minimum occurs at one of the extreme points of this set. We know by Birkhoff’s theorem that the extreme points of the set of doubly stochastic matrices are given by Permutation matrices which are all real and orthogonal. Now, we have to calculate at which point of this set of Permutation matrices do we get the minimum value $LC\sigma$. For this let us note that $\rho D$ and $\sigma D$ are diagonal matrices whose eigenvalues are arranged in ascending and descending orders respectively. As a result, global minimum occurs at $U = P = I$, such that we get that the minimum is obtained at the value $W = U\rho U\sigma^\dagger$. Now, we know that the unitary matrices which diagonalize the density matrices can be taken to be $SU(N)$ matrices. Therefore, we have performed the minimization over $SU(4)$ matrices only which gives us a better bound.
The second step of the minimization has to be made for the Bell diagonal states with the given value of the entanglement that we want to reach to. The Bell diagonal states are diagonal in the Bell basis and also we can impose an extra constraint of equal purity to that of the initial starting state for unitarily driven quantum systems. Now in the Bell basis generally the initial quantum state will not be diagonal, as a result we will need to only find the unitary operator that diagonalizes this initial density matrix. After this we find out the trace function in terms of the coefficients of the Bell diagonal state and then minimize with respect to the coefficients of the Bell diagonal state to find out the actual minimum. note that this gives the lower bound which in general will not be saturated.

\section{Connection with Lieb-Robinson bound found in many body physics or condensed matter physics}

Here, we explore the question of a relation between the concurrence speed limit derived from the quantum speed limit formalism to that of the bounds provided by the Lieb-Robinson bound formalism in the pursued mainly in the area of many body physics and condensed matter physics. Finding answers to this question can proceed along many different directions. The first direction that can be considered is a mathematical relation between these two bounds in different forms. The second direction can have a more fundamental basis, which is to say that one can explore what is the common connection between the origin of these seemingly two different kind of bounds. In this section we explore the first direction, i.e., find ordering between these two bounds in different scenarios using analytical techniques. This analysis can have implications in the time optimization of quantum information processing tasks, which we discuss in the later sections. 

For our purpose, we use the connection between the connected correlators and the concurrence for a two qubit pure state. The exact relation or equation that we use between the connected correlation function and the concurrence is that the maximum of the connected correlation function for the two qubit pure states, where the maximum is over all the operators $A$ and $B$. This relation has been proven before. Therefore, when we are looking to compare the Lieb-Robinson bound and the concurrence speed limit, we take the $t_{LRB}$ that corresponds to the maximum connected correlation function, i.e., maximized over all possible operators $A$ and $B$ in the two qubit Hilbert space. We also note here that we work with bounded Hamiltonians, in which case it suffices to optimize the distance between the states to get the speed limits. Also, we approach this from the perspective that we have different Hamiltonians as resources stores from among which we choose the ones that can optimize the time of a quantum information task. 

We also see in which cases, it is possible to derive at such an ordering and where it may not possible by this method. The conditions that we propose are those on the state parameters for which the orders between these bounds are obeyed. The question that remains open is that whether these two different kinds of bounds share a common cause with each other.

First let us consider the case where we can in principle find optimal bounds for both the cases of concurrence speed limit
obtained using quantum speed limit formalism and the Lieb-Robinson bound formalism existing in the current literature such that we can directly
compare these bounds quantitatively. First, let us consider the case of the generation of the maximal amount
of entanglement from a pure product state. The initial state can be taken as $|00\rangle$ without any loss of generality since
similar procedure can be taken for other pure states which are of the product form using the same minimization
procedure of the trace function that we have discussed before in the methods section. In this case, the set of all pure bipartite states with the entanglement value of 1 are
given by the four Bell states for the case of two qubits. The minimum of the expression $\mathrm{Tr}(\rho\sigma)$ 
is easily calculable here, and as a result we get a speed limit under a bounded Hamiltonian easily . It is straightforward to deduce what can be the value of this speed limit if we start with any type of pure bipartite product state. However, in this case it is easy to see this comparison is because both of the bounds are saturated. The next question we ask is what can we say about the other cases ? In some cases, it is not possible to find whether the Lieb-Robinson bound for the connected correlation functions is tight or not. As a result we may not be able to find the exact comparison between these bounds in those particular cases. However, the next question we ask is whether we can derive the necessary and sufficient conditions for a certain ordering of the bounds based on the state parameters? In this case, we can give some conditions for these bounds to follow a certain ordering. Also, we know that for time independent Hamiltonian, the quantum speed limit bound is saturated or optimal for some specific quantum states. Therefore we exactly know the value of the concurrence speed limit for such cases. However, for the case of arbitrary pure bipartite states of arbitrary value of entanglement, we can only have a lower bound from the Lieb-Robinson bound limit. As a result in that case, one can only give the necessary and sufficient conditions for the Lieb-Robinson bound time limit to be greater than the concurrence speed limit given by the quantum speed limit formalism. 

To be more specific, along the line of reasoning from the above paragraph, suppose we have the Lieb-Robinson speed limit as $t_{LRB}\geq  c$ and the concurrence speed limit is saturated or tight. Then, in that case, we can find the conditions such that the $t_{LRB} \geq c \geq  t_{CSL}$ in terms of the state parameters. Mathematically, from the Lieb-Robinson bound we have the following
\begin{align}
t_{LRB}\geq \frac{\ln|u_2(A_1A_2)|-\ln c_2+R}{v_{LR}}.
\end{align}
Now, suppose we have a bipartite pure state of two qubits as $|\Psi_{AB}\rangle $ with a given value of concurrence, that has been evolved
from a product state $|\Psi_{A}\rangle|\Psi_{B}\rangle $ by a non-local time-independent Hamiltonian with short range interactions. In this
case the concurrence speed limit is given by the following
\begin{align}
t_{CSL}=\cos^{-1}\frac{|\langle\Psi|\Phi\rangle}{\omega}
\end{align}
From the above equations, we can see that in this case, we can only give conditions on the state parameters such that
the $t_{CSL}$ is smaller than the $t_{LRB}$. From the above equations we also get the following
\begin{align}\nonumber
\ln|u_2(A_1A_2)|=\ln|\mathrm{Tr}(A_1A_2\rho)-\mathrm{Tr}(A_1\rho_1)\mathrm{Tr}(A_2\rho_2)|,
\end{align}
which is just the expression for the natural logarithm of the connected correlation function. The base of the logarithm can be fixed according to the need. Now, we would like to find conditions on the state parameters of $|\Psi\rangle, |\Phi\rangle $ such that the following equation holds
\begin{align}\nonumber
\frac{\ln|\mathrm{Tr}(A_1A_2\rho)-\mathrm{Tr}(A_1\rho_1)\mathrm{Tr}(A_2\rho_2)|-\ln c_2+R}{v_{LR}}\\ \nonumber \geq \cos^{-1}\frac{|\langle\Psi|\Phi\rangle}{\omega},
\end{align}
which can reveal certain ordering of the two bounds in the existing literature as proposed in the section above. Now, as a progression from this to more research lying at the intersection of quantum speed limit and  Lieb-Robinson bound, more development in the optimality conditions, tightness and saturation conditions of these bounds will lead us to have more quantitative and tighter conditions on the ordering of these bounds that can become useful in resource determination strategies in the field of quantum information science and application of many body physics tools in the field of quantum science and emerging quantum technology.

On another note, not only concurrence, but localizable entanglement has also been found to have a strict inequality relation with the connected correlation function, not only in bipartite case, but also in multipartite cases. As a result, this has the potential to develop into a research area that can explore the speed limits in many body physics using multipartite connected correlation function and the localizable entanglement in quantum information science.

\section{Conclusions and Future Directions}

In this article we addressed the question `what is the minimum time to generate a certain amount of quantum entanglement starting from an arbitrary initial quantum state'. We found bounds for this question for the concurrence for arbitrary mixed quantum states for two qubits. We have found two different ways this question can be solved providing with exact and rigorous results. These results have immediate application in the quantum information and quantum computation platform. This is because quantum entanglement is a valuable resource that is used by various quantum information and computation task and quantum communication tasks. Therefore in those cases our analysis will provide answers and directions to the method of preparing a benchmark for attainment of a certain amount of quantum entanglement in a time efficient way. Also we make a connection with the Lieb-Robinson bound that has applications in many body physics. Therefore our results hold the potential to be useful in the area of quantum information, quantum computation and also many body physics. As future directions, one area would be to have a more rigorous relationship between the quantum speed limit and the Lieb-Robinson bound. Another promising direction would be to extend the answers to this particular question of generating a given amount of entanglement to the case of general mixed quantum states for the case of qubits and genuine multipartite entanglement or multipartite entanglement in general which may involve the use of localizable entanglement.

\begin{appendix}
    \begin{figure}
    \includegraphics[scale=1]{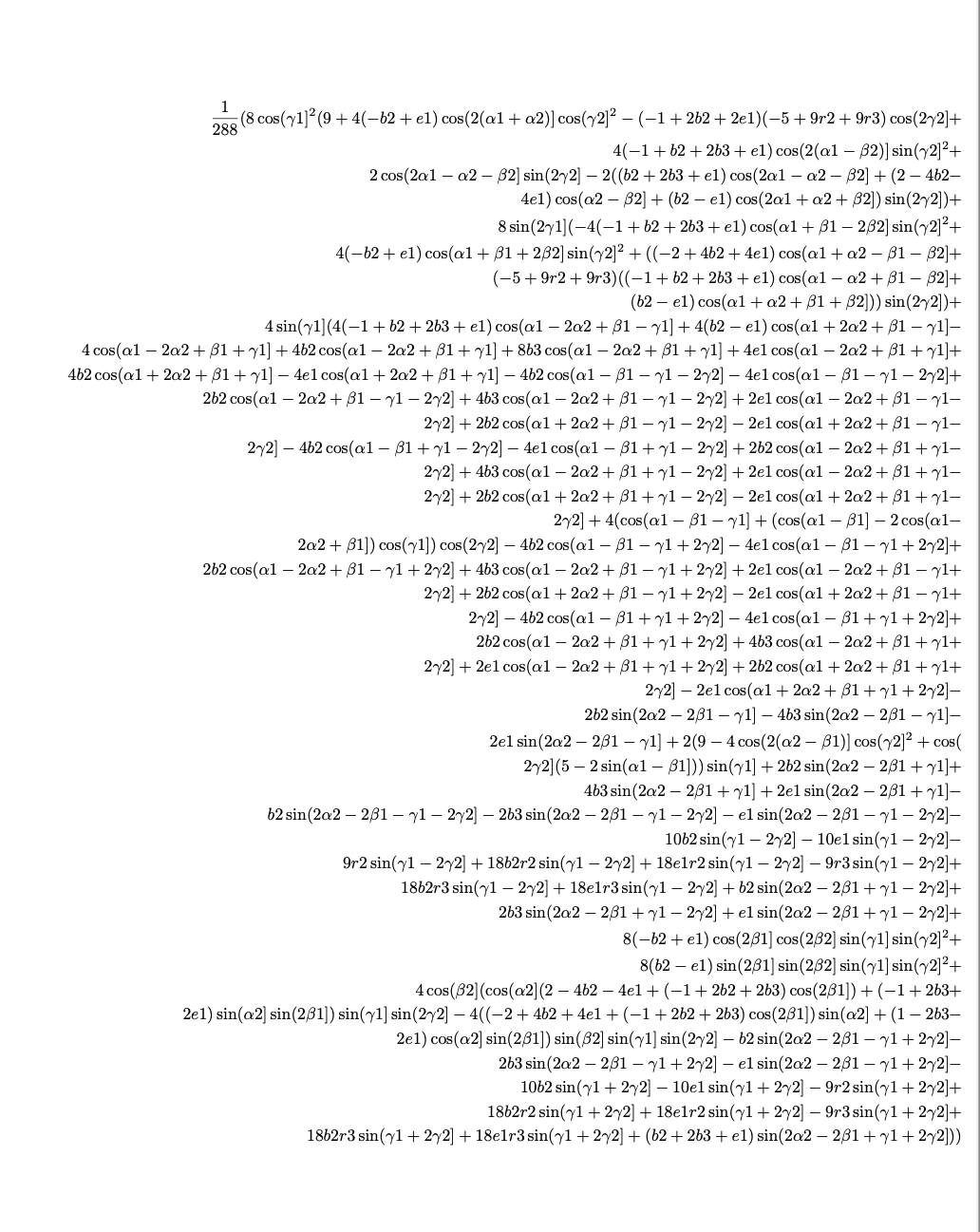}
    \end{figure}
\end{appendix}

\end{document}